# High–order correlations of rich galaxy clusters

Alberto Cappi[1] and Sophie Maurogordato[2]

[1] Osservatorio Astronomico di Bologna, via Zamboni 33, I–40126 Bologna, Italy

[2] CNRS; Observatoire de Paris–Meudon, LAM, 5 Place J.Janssen, F–92195 Meudon, France

E–MAIL:

A. CAPPI: cappi@astbo3.bo.astro.it

S. MAUROGORDATO: maurogordato@melamc.obspm.circe.fr

# Abstract

We analyse the two–dimensional all–sky distribution of rich Abell and ACO galaxy clusters by using counts in cells and measuring the high–order area–averaged angular correlation functions. Confirming previous results, we find a well defined hierarchical relation between the two and three–point correlation functions, remarkably constant with scale. In the angular range $2° \leq \theta \leq 4°$, the southern sample, limited at $b_{II} \leq -40°$ and including both Abell and ACO clusters, shows a remarkable hierarchical behavior up to the sixth order, while northern Abell clusters give positive correlations in the same range only up to the fourth order.

The inferred deprojected values of the 3–D coefficients $S_J$, where $S_J = \bar{\xi}_J/\bar{\xi}_2^{J-1}$, are similar to those measured for the galaxy distribution, and consistent with theoretical predictions. These results are confirmed to the 4th order by our analysis of a 3–D sample of Abell and ACO clusters. Assuming that selection effects and / or the absence of a cluster fair sample are the reason of the difference between the two galactic hemispheres, and between Abell and ACO clusters, our results indicate that the statistical properties of the cluster distribution originate from the underlying galaxy distribution and show that the biasing between clusters and galaxies is non–linear.

*Subject headings:* Galaxies: clustering — large-scale structure of the universe

# 1  Introduction

The study of the large-scale structure in the Universe is a central issue in cosmology. A major challenge for theories of galaxy formation is to explain the observed properties of the galaxy distribution. In this field, a statistical approach is natural and fruitful (see Peebles 1980); this means, of course, that large, well defined samples are needed.

Rich clusters represent the highest peaks of the galaxy distribution, and they are potentially a powerful tool for the analysis of the large-scale structure. Their use as tracers of the luminous matter distribution has been possible after the publication of the Abell catalog of clusters (Abell 1958), whose extension by Abell, Corwin & Olowin (1989) gives a whole-sky coverage except at low galactic latitude.

The spatial two-point correlation function $\xi_{cc}(r)$ of Abell clusters was measured by Bahcall & Soneira (1983) and Klypin & Kopylov (1983), after a first estimate of the angular two-point correlation function $\omega_{cc}(\theta)$ by Hauser & Peebles (1973): they found that it can be fitted with a power-law comparable to that of galaxies with slope $\gamma \sim -1.8$, but with a much larger correlation length. In order to explain the different amplitudes of the galaxy and cluster autocorrelations, Kaiser (1984) and Bardeen et al. (1986) introduced the concept of bias: the selection of rare peaks above some fixed threshold in a smoothed Gaussian density field amplifies the correlation function of such peaks relatively to the underlying distribution. Nevertheless, this mechanism is not sufficient to reconcile the standard CDM scenario with the statistical properties of clusters (Bahcall & Cen 1992; Olivier et al. 1993). A better description of a distribution requires the knowledge of the higher-order correlation functions. The three-point correlation function $\xi_3$ is particularly interesting: in the biased galaxy formation scenario, for such dense systems as clusters, its decomposition should give a term of order 3 (Politzer & Wise 1984) which is not observed. The 2-D analysis of Jing & Zhang (1989) and Tóth, Hollósi & Szalay (1989), and the 3-D analysis of Jing & Valdarnini (1991), have shown that the following hierarchical relation holds for Abell and ACO

(Abell et al. 1989) clusters:

$$\xi_3 = Q[\xi_2(r_a)\xi_2(r_b) + \xi_2(r_b)\xi_2(r_c) + \xi_2(r_a)\xi_2(r_c)] \tag{1}$$

with $0.6 \leq Q \leq 1.1$. This is the same relation previously found for galaxies (Groth & Peebles 1977; Davis & Peebles 1977); its generalization at higher orders gives the so-called factorized hierarchical models (see Fry, 1984a & b; Schaeffer 1984), where the correlation functions at all orders can be written in the following way:

$$\xi_J(\mathbf{r}_1, ..., \mathbf{r}_J) = \sum_\alpha Q_J^{(\alpha)} \sum_{ij} \Pi^{N-1} \xi_2(r_{ij}) \tag{2}$$

Each term of equation (2) is associated to a graph, points $\mathbf{r}_J$ correspond to nodes and factors $\xi_2(r_{ij})$ correspond to edges between nodes $i$ and $j$. Such a hierarchy has a tree structure, and each distinct tree shape (or topologically distinct graph) is denoted by $\alpha$, while the sum over $ij$ denotes a sum over relabellings of a given tree $\alpha$. This relation implies an equivalent one in 2-D, where the coefficients $Q_J^\alpha$ are changed into the projected $q_J^\alpha$:

$$\omega_J(\theta_1, ..., \theta_J) = \sum_\alpha q_J^{(\alpha)} \sum_{ij} \Pi^{N-1} \omega_2(\theta_{ij}) \tag{3}$$

In 2-D, the relation was shown to hold on galaxy catalogs up to $J = 4$ (Fry & Peebles 1978; Sharp et al. 1984). Szapudi, Szalay & Boschán (1992) found that galaxies in the Lick catalog continue to follow the hierarchy up to $J = 8$, and confirmed the validity of the relation (1) for clusters selected in the same catalog (see also Borgani, Jing & Plionis 1992).

Hierarchical models represent a subset of the so-called scale-invariant (SI) models (Balian & Schaeffer 1989), defined by the scaling property:

$$\xi_J(\lambda r_1, ..., \lambda r_J) = \lambda^{-(J-1)\gamma} \xi_J(r_1, .., r_J) \tag{4}$$

This property is verified by a solution of the BBGKY equations for $\Omega \leq 1$.

The void probability function (VPF), which depends on all the $J$–point correlation functions (White 1979), allowed to confirm the scale–invariance for galaxies (Bouchet & Lachièze-Rey 1986; Maurogordato & Lachièze-Rey 1987) and clusters (Jing 1990; Cappi, Maurogordato, & Lachièze-Rey 1991). However, the VPF gives global information. In a comparison with theoretical predictions, derived from perturbative calculations or N–body simulations, we are interested also in a direct knowledge of each high–order moment. In SI models, the volume–averaged correlation functions, or connected moments, $\bar{\xi}_J = V^{-J} \int_V \xi_J d^3r_1...d^3r_J$, (we will assume that the average is made over a sphere of radius $R$) can be written as:

$$\bar{\xi}_J = S_J \bar{\xi}_2^{J-1} \tag{5}$$

where the $S_J$ are independent of $R$, and can be expressed as an integral over the unit volume: $S_J = \int_V d^3\rho_1, ..., d^3\rho_J \xi_J(r_0\rho_1, ...r_0\rho_J)$. In the special case of the hierarchical models, and assuming that the $Q_{J,\alpha}$ are independent of $\alpha$, then:

$$\xi_J(\mathbf{r}_1, ..., \mathbf{r}_J) = Q_J \sum_\alpha \sum_{ab} \Pi^{N-1} \xi_2(r_{ab}) \tag{6}$$

This means that the $S_J$ and the $Q_J$ coefficients can be linked by the following relation: $S_J = F_J J^{J-2} Q_J$, where the coefficients $F_J$ depend on the shape of the integration volume and on the slope of the two–point correlation function. At order 3,

$$S_3 = 3Q_3 \frac{J_3}{J_2^2} \tag{7}$$

where $J_2 = 72/(3-\gamma)(4-\gamma)(6-\gamma)2^\gamma$ and $J_3 = (3/4\pi)^3 \int_0^1 d^3x_1 d^3x_2 d^3x_3 (\|x_1 - x_2\| \|x_2 - x_3\|)^{-\gamma}$ (Peebles 1980). Recently Gaztañaga & Yokohama (1993) have published the analytical solution of the $J_3$ integral assuming a power-law two-point correlation function, and at higher orders numerical estimates are available (Gaztañaga 1994). Bouchet et al. (1993) have shown that IRAS galaxies follow the hierarchical relation (5) up to order 5, and similarly Gaztañaga (1992) has found that also CfA and SSRS galaxies follow the above relation up to $J = 4$.

In this paper we present estimates of the area–averaged correlation functions for the 2–D cluster distribution and we check them using also a 3–D sample, with the aim to test directly the validity of the scale–invariance for galaxy clusters, and to compare the high–order moments with those measured for the galaxy distribution and expected from the theory. The method is presented in section 2; section 3 is devoted to the definition of the samples and the presentation of the main results; in section 4 the implications of these results are discussed. Our main conclusions are summarized in section 5.

## 2 Counts in cells and moments

We calculate the moments of the cluster distribution from counts in cells (see Bouchet, Davis & Strauss 1992, Gaztañaga 1992, and Bouchet et al. 1993). Here we describe the procedure for the analysis of a 2–D catalog. Given a sample of objects with known angular coordinates, we place randomly a fixed number of circles with radius $\theta$ and area $2\pi[1 - \cos(\theta)]$. A number of circles $N_{ran} = 10^6$ was found sufficient to avoid undersampling at small scales and to have negligible fluctuations. Circles which cross the sample borders are not counted. We measure the count probabilities $P(N)$ of finding $N$ objects within a radius $\theta$, by counting the number of circles containing $N$ objects and dividing it by the total number of circles. This procedure is repeated at different scales $\theta$.

The same method can be applied to the analysis of 3–D volume–limited samples; instead of circular areas, spherical volumes of radius $R$ are used. From the $P(N)$ we can estimate the centered moments of order $J$ (see Peebles 1980):

$$\mu_J = \sum_N P(N) \left(\frac{N - \bar{N}}{\bar{N}}\right)^J \qquad (8)$$

where $\bar{N} = \sum N P(N)$, i.e. at each scale we compute directly the effective $\bar{N}$ from counts. This means that the effective $\bar{N}$ may slightly differ from the sample $\bar{N} = nV$, where $n$ is the mean density of the sample. We choosed to use the effective $\bar{N}$ for self–consistency; anyway we verified that the two methods of computing $\bar{N}$ give similar results.

The area (or volume) –averaged correlation functions (the reduced moments) are finally given by the following relations, where the Poisson shot noise is subtracted from each moment:

$$\bar{\omega}_2 = \mu_2 - \frac{1}{\bar{N}} \tag{9}$$

$$\bar{\omega}_3 = \mu_3 - 3\frac{\mu_2}{\bar{N}} + \frac{2}{\bar{N}^2} \tag{10}$$

$$\bar{\omega}_4 = \mu_4 - 6\frac{\mu_3}{\bar{N}} + 11\frac{\mu_2}{\bar{N}^2} - 3\mu_2^2 - \frac{6}{\bar{N}^3} \tag{11}$$

$$\bar{\omega}_5 = \mu_5 - 10\frac{\mu_4}{\bar{N}} + 35\frac{\mu_3}{\bar{N}^2} - 50\frac{\mu_2}{\bar{N}^3} + 30\frac{\mu_2^2}{\bar{N}} - 10\mu_2\mu_3 + \frac{24}{\bar{N}^4} \tag{12}$$

$$\bar{\omega}_6 = \mu_6 - 15\frac{\mu_5}{\bar{N}} + 85\frac{\mu_4}{\bar{N}^2} - 225\frac{\mu_3}{\bar{N}^3} + 274\frac{\mu_2}{\bar{N}^4} - 255\frac{\mu_2^2}{\bar{N}^2} + 30\mu_2^3 - 10\mu_3^2 + 150\frac{\mu_2\mu_3}{\bar{N}} - 15\mu_2\mu_4 - \frac{120}{\bar{N}^5} \tag{13}$$

## 3 Analysis of the samples

### 3.1 Definition of the 2–D samples

We analysed both 2–D and 3–D cluster samples. The advantage of using 2–D samples is the larger number of objects, which allows a better estimate of the higher moments; but in this way one can only indirectly obtain the corresponding 3–D values, and for this reason the analysis of smaller 3–D samples is a complementary and useful check.

The 2–D sample includes 1674 clusters (NS45D61), 774 Abell clusters in the north (N45D61) and 900 Abell + ACO clusters in the south (S45D61) –hereafter for Abell clusters we will mean the clusters at $\delta \geq -27°$ selected by Abell (1958), and for ACO clusters the southern clusters selected by Abell, Corwin, Olowin (1989)– with $\mid b_{II} \mid \geq 45°$, richness class $R \geq 1$ and distance class $D \leq 6$, covering an area of 3.68 steradians (see table 1). The high cut in galactic latitude allows to minimize the effect of galactic extinction (we analysed also a sample limited at $\mid b_{II}\mid\geq 40°$, without finding significantly different results).

For what concerns projection effects, Olivier et al. (1990) found that projection contamination affects significantly $\omega_{cc}(\theta)$ only at scales below 1°. Peculiar motions of clusters are small relatively to their Hubble velocities (Huchra et al., 1990), and distortions in redshift space should be negligible, as well as their effect on the estimates of $S_J$. Finally, the southern samples are a mixture of Abell and ACO clusters; while having been selected according to the same criteria, they show some differences (see Scaramella et al. 1991) which nonetheless might be partly real (e.g. Cappi & Maurogordato 1992).

Errors were computed by the bootstrap technique (see Ling, Frenk & Barrow 1986). We generated 50 catalogs from each sample, then we estimated errors in two ways. For each catalog, we found the values of $s_J$ and $\gamma_J$ from a linear fit of the relation $\log \bar{\omega}_J = A + \gamma_J \log \bar{\omega}_2$ in the range $2° \leq \theta \leq 4°$; then we estimated the respective rms. We also fixed $\gamma_J = J - 2$, and we estimated the rms of $s_J$ at different angular scales.

## 3.2 Analysis of the 2–D samples

Figures 1a–d show $\bar{\omega}_J$, $J = 3,..,6$, as a function of $\bar{\omega}_2$. Looking first at the total sample (black hexagons), a very good hierarchical behavior appears, especially in the range $2° \leq \theta \leq 4°$. It is noteworthy to stress that the straight lines in the figures *are not* fits of the points: they simply represent the relation $\bar{\omega}_J = J^{J-2}\bar{\omega}_2^{J-1}$, which differs from the hierarchical one by a factor of the order of unity (e.g. Gaztañaga 1994). Table 2 lists the average values of $s_J$ calculated assuming the hierarchical relation (5) between 2° and 4°, and also the values of $\gamma_J$ and $s_J$ computed from a least–square fit to the relation $\log(\bar{\omega}_J) = \log(s_J) + \gamma_J \log(\bar{\omega}_2)$, leaving the slope $\gamma_J$ as a free parameter, with the corresponding bootstrap errors. Notice that for the fit we kept the range fixed ($2° \leq \theta \leq 4°$) at all orders.

Figures 2 and 3 show respectively $\bar{\omega}_J$ and $s_J$ as a function of scale. The correlations are consistent with power–laws systematically steepening at higher orders. $s_3$, $s_4$, $s_5$ and $s_6$ are nearly constant between $2°$ and $4°$, corresponding roughly to a spatial range of 20–40 $h^{-1}$ Mpc. The analysis is limited at small scales by shot noise, and at large scales by the small number of independent areas, with an increasing error in the estimate of $P(N)$ at large $N$, and by the loss of the $P(N)$ exponential tail, with a systematic underestimate of the higher moments. Moreover, at large scales systematic discrepancies appear between Abell and ACO clusters (see Olivier et al. 1990).

Up to now, we have discussed the total sample, stressing its hierarchical behavior. But the situation appears more complex when looking separately at the northern and southern samples, because a systematic discrepancy is apparent. In figures 1a–d, it is possible to see that most of the hierarchical behavior is due to the southern sample. Northern Abell clusters follow the hierarchical behavior at order 3 (with a lower $s_3$ than southern clusters), but they deviate from the expected behavior at the 4th order, and their $\bar{\omega}_5$ becomes negative very rapidly, while the S45D61 sample shows $s_J$ nearly constant in the range $2° \leq \theta \leq 4°$ up to the 6th order. These effects are apparent in figure 4, which shows $s_J$ as a function of $\theta$ for the northern and southern samples separately.

That a difference could be detected is not surprising. For example, $\bar{\omega}_2(\theta)$ of southern clusters ($b_{II} \leq -45°$) is lower than the northern one at large angles (as found for the standard 2–point correlations by Olivier et al. 1990), while in the range $2°$–$4°$ the values of $s_J$ for southern clusters are systematically higher than the northern ones. These effects become more important not only at larger $\theta$, but also at higher orders.

Such systematic differences constitute the main limit of any inference from the Abell and ACO cluster distribution.

In principle, according to bootstrap errors, the difference between the northern and southern $s_J$ at high orders is significant, but errors derived with the bootstrap technique, while more realistic than Poissonian ones, might be an underestimate.

Moreover, it is probable that we do not dispose yet of fair samples. The presence of some large structures might change drastically the estimated count probabilities, especially at the tail of the distribution. This might explain the relative insensitivity of the two-point correlation function or the VPF to these differences, which become increasingly large at higher orders.

Also selection effects in the Abell and ACO cluster catalog might cause some differences, and the ACO catalog might be deeper than the Abell one (see e.g. Plionis et al. 1992). We checked this possibility analysing two subsamples, the first made only of Abell clusters at southern galactic latitude ($b_{II} \leq -40°$, $\delta \geq -17°$), the second made only of ACO clusters ($b_{II} \leq -40°$, $\delta \leq -27°$). The results are quite noisy, especially at the higher orders, nevertheless they show a systematic difference between ACO and Abell clusters, and only ACO clusters show clear signs of a hierarchical behavior at orders $> 4$.

We suspect that a combination of the above discussed effects is at the origin of this discrepancy. Nevertheless, it is interesting to note that the combination of these subsamples gives a *stronger* hierarchical behavior; this fact gives us some confidence that cluster correlations are indeed hierarchical, but that in present catalogs this hierarchy appears clearly only at the lowest orders, where the signal-to-noise is sufficiently high.

## 3.3 Deprojection of $s_J$

We have obtained the projected $s_J$, and the next step is the estimate of the 3-D $S_J$.

The values found for the total sample NS45D61 –assuming that they represent our best estimates of the correlation hierarchy (5) at scales between 2° and 4° degrees, corresponding approximately to the range $20 - 40$ $h^{-1}$ Mpc– are the following: $s_3 \sim 3.13$, $s_4 \sim 16$, $s_5 \sim 110$, $s_6 \sim 745$. The derivation of $Q_J$ from $s_J$ requires two steps. Assuming that the the spatial correlation functions $\bar{\xi}_J$ follow the hierarchical relation (5), a 2–D projection gives the corresponding relation for the angular correlation functions $\bar{\omega}_J$, where the 2–D $q_J$ differ from the 3–D $Q_J$ by a multiplicative factor $r_J$, $q_J = r_J Q_J$:

$$\sigma^3 q_J^{(\alpha)} \sum_{ij} \Pi^{N-1} \omega_2(\theta_{ij}) = \rho^3 Q_J^{(\alpha)} \int_0^\infty r_1^2 dr_1 \Phi(r_1) \int_0^\infty r_2^2 dr_2 \Phi(r_2)$$
$$... \int_0^\infty r_J^2 dr_J \Phi(r_J) \sum_{ij} \Pi^{N-1} \xi_2(r_{ij}) \qquad (14)$$

In equation (14) $\Phi(r)$ is the selection function of the sample, $\sigma$ and $\rho$ are respectively the 2–D and 3–D densities. This equation can be solved analytically for small separations, assuming a power-law for the two-point correlation function.

Tóth et al. (1989), assuming a step–like form for the selection function and a power-law form for the two–point correlation function, have calculated analytically the ratio between $Q_3$ and $q_3$, finding $Q_3 = q_3/1.040$. Bernardeau (1994, private communication) has extended the calculation up to the 6th order. At this point, $\omega_J$ can be computed from the area–averaged $\bar{\omega}_J$. At each order, this implies the computation of an $f_J$ integral and a sum over all possible vertices, leading to $s_J = f_J J^{J-2} q_J$. Numerical estimates of the $f_J$ integrals (Bernardeau 1994, private communication) show that the corrective factors are around unity, and not very sensitive to the geometry or to variations of the $\xi_2$ slope. In the same way, the 3–D $S_J$ and $Q_J$ are related through $S_J = F_J J^{J-2} Q_J$, where the $F_J$ are integrations over the 3–D box. $F_J$ have been calculated analytically at order 3 (Gatzanaga and Yokohama 1993) and numerically at higher orders (Gatzanaga 1994) in the case of a power-law slope with $\gamma = -1.8$ for the correlation function. In this way we can obtain a first rough estimate of the values of $S_J$ deduced from the 2–D analysis. Table 3 gives the main quantities involved in this computation, as well as the values of $Q_J$ and $S_J$. The estimated values of $Q_J$ for clusters are $Q_3 = 1.0$, $Q_4 = 0.9$, $Q_5 = 0.7$, $Q_6 = 0.4$. Detailed calculations of these factors, as well as a more complete analysis using a selection function computed from the data will be presented in another paper (Bernardeau, Cappi, & Maurogordato

1994).

## 3.4 Analysis of the 3–D sample

We checked our results analysing a nearly complete spatial sample limited at $\mid b_{II} \mid \geq 40°$ and $z \leq 0.08$, with 281 clusters (120 Abell clusters in the north and 161 Abell + ACO clusters in the south, including richness class 0). It is an extension of the sample defined by Cappi & Maurogordato (1992), without the limit on the apparent magnitude of the tenth cluster member ($m \leq 16.4$) derived from the Postman et al. sample (1992). Comoving distances in Mpc are calculated from redshifts using the Mattig formula:

$$D = \frac{c}{H_0}\frac{1}{q_0^2}[1 - q_0 + q_0 z + (q_0 - 1)\sqrt{2q_0 z + 1}]/(1 + z) \qquad (15)$$

where we fixed $H_0 = 100$ and $q_0 = 0.5$ (the only choice which allows using euclidean geometry for separations and volumes).

The 3–D sample is too small to give significant results for the moments higher than 4, and we cannot test if the discrepancy found in the analysis of the 2–D samples is reproduced here. Considering the total sample (both hemispheres), we obtain $S_3 \sim 2.5$ between 22 and 38 h$^{-1}$ Mpc, and $S_4 \sim 12$ between 30 and 35 h$^{-1}$ Mpc (fig.5).

Here again, we find a systematic difference between the northern and southern samples; in particular, we cannot measure a positive $\bar{\xi}_4$ for the northern sample, except for a small range around $\theta \sim 2$ degrees. It may be interesting to note that the northern and southern samples are different also in their percolation properties, and that this can be explained by the presence of the large Horologium–Reticulum supercluster (Cappi & Maurogordato 1992). Therefore it is possible that the systematic difference, which becomes significant at high orders (see also the multifractal analysis of Borgani et al., 1994), is partly due to a real difference in the cluster distribution. The 2–D and the 3–D analysis give consistent values of $S_J$ and comparable scales where the hierarchical behavior is observed. This is noteworthy, because the 3–D sample is limited at $z \leq 0.08$ and includes richness 0 clusters, while the 2–D samples are mainly composed by clusters of distance class 5 and 6 and richness class $R \geq 1$. We can conclude that clusters follow a hierarchical behavior up to the 4th order, and –from the analysis of the 2–D samples– that southern clusters follow a hierarchical behaviour up to the 6th order.

## 4 Discussion

Recovering the 3–D coefficients $Q_J$ and $S_J$ allows a comparison between our results and those obtained for different galaxy samples, and also with theoretical predictions. From our 2–D analysis we find $Q_3 \sim 0.98$, comparable to the values found with other methods. Gaztañaga (1994) estimated $S_J$ –up to the 9th order– from the galaxy angular distribution in the APM Survey. At scales between 7 and 30 h$^{-1}$ Mpc, he finds $S_3 = 3.1 \pm 0.14$, $S_4 = 22.0 \pm 2.5$, $S_5 = 146 \pm 47$, and $S_6 = 1514 \pm 337$. A comparison of these values with the corresponding cluster values reported in table 3 shows that the galaxy and cluster $S_3$ values are remarkably similar, while their $S_5$ are consistent at the 1–$\sigma$ level, their $S_4$ values differ at the 2–$\sigma$ level, and only $S_6$ values are different at the 3–$\sigma$ level. While uncertainties (depending also on the deprojection factors) do not allow a detailed comparison, these results suggest that the APM Galaxy Survey and rich clusters are sampling the same underlying large–scale distribution. In the hierarchical tree models studied by Bernardeau and Schaeffer (1992), for high density objects such as galaxies or clusters, the values of all $Q_J(x)$ converge to 1 for large $x$ –and even as soon as $x > 1$– (where $x = N/N_c$ is the scale parameter). The values of $Q_J$ for clusters are in good agreement with these models, except for $Q_6$, which however has the highest error, as clearly indicated by the large bootstrap errors on $s_6$ and $\gamma_6$. The values of $S_J$ are also similar to the predictions derived in the frame of perturbation theory; assuming a self similar power spectrum and a spectral index $n = -1$, one gets $S_3 = 2.86$ (Juskiewicz, Bouchet, Colombi 1993; Bernardeau 1994) and $S_4 = 23$ (Bernardeau 1994; Catelan & Moscardini 1994); from numerical simulations with scale–free Gaussian initial conditions, Lucchin et al. (1994) find (again for $n = -1$) $S_3 \simeq 3.1 - 3.3$, $S_4 \simeq 16 - 18$ and $S_5 \simeq 135 - 147$, while in their N–body simulations of a Cold+Hot Dark Matter universe Bonometto et al. (1994) find $S_3 \simeq 2.5$ and $S_4 \simeq 7.5$ for galaxies.

Following Fry and Gazañaga (1993), the smoothed biased (galaxy) density field can be expressed as a Taylor series of the matter density field:

$$\delta_b = \sum_{k=0}^{\infty} \frac{b_k}{k!} \delta_M^k \qquad (16)$$

In the linear approximation, we have $\delta_b \sim b\delta_M$ and $\xi_{2b} \sim b^2 \xi_{2M}$. It follows that the "normalized" skewness $S_3$ should be inversely proportional to the bias:

$$S_{3b} = <(b\delta_M)^3> / <(b\delta_M)^2>^2 = S_{3M}/b \qquad (17)$$

If we take into account higher orders in $\delta_M$ in the development of $\delta_b$, the relation between $\xi_J$ and $\xi_2$ is affected. For instance, taking only the leading term in $\xi_{2M}$, we have for $S_3$:

$$S_{3b} = b^{-1}(S_{3M} + 3b_2/b) \qquad (18)$$

while new terms appear at higher $J$ (Fry & Gaztañaga 1993; Bernardeau 1994).

For $J = 2$, the linear bias expression for $\xi_2$ coincides with the leading term in the non–linear bias expression, and we have $\xi_{2b} \sim b^2 \xi_{2M}$ in both cases. From this relation, it is possible to measure for example the relative bias between clusters and galaxies (see Peacock & Dodds 1994).

For $J = 3$, the linear biasing –equation (17)– and the non–linear biasing –equation (18)– give two different relations, and this allows to discriminate between the two cases.

We have shown that galaxies and clusters have nearly the same $S_3$: as a consequence, a linear biasing is ruled out.

Moreover, the fact that the values of $Q_J$ are around unity for both galaxies and clusters is consistent with models where a natural bias comes out from gravitational interaction of the matter field.

## 5  Conclusions

We have calculated the high–order connected moments (up to order 6) of the rich cluster distribution using counts in cells. Our main results are based on the analysis of the 2–D distribution, and on a smaller 3–D sample.

We have shown that the all–sky 2–D and 3–D cluster distribution is consistent with the correlation hierarchy (5) up to $J = 4$, and the 2–D southern cluster distribution up to $J = 6$, at scales between 2° and 4° degrees, corresponding approximately to the range 20 – 40 $h^{-1}$ Mpc.

The inferred values for $S_J$ are comparable to those found for galaxy samples; in particular, the similar values of $S_3$ for galaxies and clusters imply a non–linear bias.

Anyway, subsamples show systematic differences. Northern Abell clusters show a clear hierarchical behavior only at order 3. These differences can be due to a combination of selection effects, smallness of samples and differences in the real cluster distribution. The fact that the combined, largest sample shows a well defined hierarchical behavior even at the 6th order, suggests that cluster correlations are hierarchical, but only larger samples will allow to confirm this result.

These statistical properties may give stronger constraints to the various models of galaxy formation even if the accurate definition of clusters in N–body simulations on one hand, and the construction of unbiased, fair cluster samples on the other hand, represent problems yet to be solved.

**Acknowledgements**

We wish to thank Francis Bernardeau for his critical reading of the manuscript, François Bouchet and Richard Schaeffer for useful discussions, and the referee, Joel

# Figure captions

**Fig. 1 a–d** Area–averaged angular correlations $\bar{\omega}_J$ ($J = 3, 4, 5, 6$) as a function of the two–point correlation $\bar{\omega}_2$. for the total sample NS45D61 (black hexagons), the northern sample N45D61 (open squares) and the southern sample S45D61 (open hexagons). Solid lines are power–laws with the following parameters: a) $s_3 = 3$, $\gamma_3 = 2$; b) $s_4 = 16$, $\gamma_4 = 3$; c) $s_5 = 125$, $\gamma_5 = 4$; d) $s_6 = 1296$, $\gamma_6 = 5$.

**Fig. 2** Area–averaged angular correlation functions $\bar{\omega}_J$, $J = 2, 3, 4, 5, 6$, as a function of scale $\theta$, for the total sample NS45D61. Lines represent least–squares fit of the points, $\log(\bar{\omega}_J) = A_J + B_J \log(\theta)$, within the range $2° \leq \theta \leq 4°$. Best–fit parameters are the following: $A_2 = 0.033$, $B_2 = -0.95$; $A_3 = 0.538$, $B_3 = -1.84$; $A_4 = 1.275$, $B_4 = -2.79$; $A_5 = 2.183$, $B_5 = -3.81$; $A_6 = 3.093$, $B_6 = -4.86$.

**Fig. 3** Values of $s_J = \bar{\omega}_J / \bar{\omega}_2^{J-1}$ ($J = 3, 4, 5, 6$) for the total sample NS45D61. Bootstrap errors are shown in the range $1° \leq \theta \leq 5°$, with step equal to $1°$.

**Fig. 4** Values of $s_J = \bar{\omega}_J / \bar{\omega}_2^{J-1}$ for the S45D61 (open points) and N45D61 (black points) samples. $J = 3$: triangles, $J = 4$: squares, $J = 5$: pentagons, $J = 6$: hexagons. Bootstrap errors are shown in the range $1° \leq \theta \leq 5°$, with step equal to $1°$.

**Fig. 5** Values of $S_J = \bar{\xi}_J / \bar{\xi}_2^{J-1}$ ($J = 3, 4$), resulting from the analysis of the 3–D cluster sample. Bootstrap errors are shown in the range 20–40 $h^{-1}$ Mpc, with step equal to 5 $h^{-1}$ Mpc.

Table 1: Definition of the 2–D samples.

|  | N45D61 | S45D61 |
|---|---|---|
| $b_{II}$ | $+45°$ | $-45°$ |
| $N_c$ | 774 | 900 |
| $R_{min}$ | 1 | 1 |
| $D_{max}$ | 6 | 6 |
| $z_{max}$ | — | — |

Table 2: Average values and rms of $s_J$ between $2°$ and $4°$, and values of $s_J$ and $\gamma_J$ estimated from a linear fit of the relation $\log \bar{\omega}_J$–$\log \bar{\omega}_2$ in the same range, with the corresponding bootstrap errors.

| J | 3 | 4 | 5 | 6 |
|---|---|---|---|---|
| $s_J$ (mean and rms) | $3.13 \pm 0.06$ | $16 \pm 1$ | $110 \pm 12$ | $745 \pm 123$ |
| $s_J$ (fit) | $2.98^{+0.25}_{-0.22}$ | $15.2^{+2.4}_{-1.7}$ | $114^{+33}_{-13}$ | $855^{+196}_{-29}$ |
| $\gamma_J$ | $1.95 \pm 0.09$ | $2.96 \pm 0.34$ | $4.04 \pm 1.37$ | $5.16 \pm 3.84$ |

Table 3: Factors used for deprojecting the 2–D estimates of $s_J$.

| J | 3 | 4 | 5 | 6 |
|---|---|---|---|---|
| $s_J$ | 3.13 | 16 | 110 | 745 |
| $f_J$ | 1.02 | 1.03 | 1.08 | 1.15 |
| $F_J$ | 1.03 | 1.10 | 1.15 | 1.25 |
| $r_J$ | 1.04 | 1.11 | 1.22 | 1.36 |
| $Q_J$ | 0.98 | 0.87 | 0.67 | 0.37 |
| $S_J$ | 3.04 | 15.4 | 96 | 595 |

Figure 1a

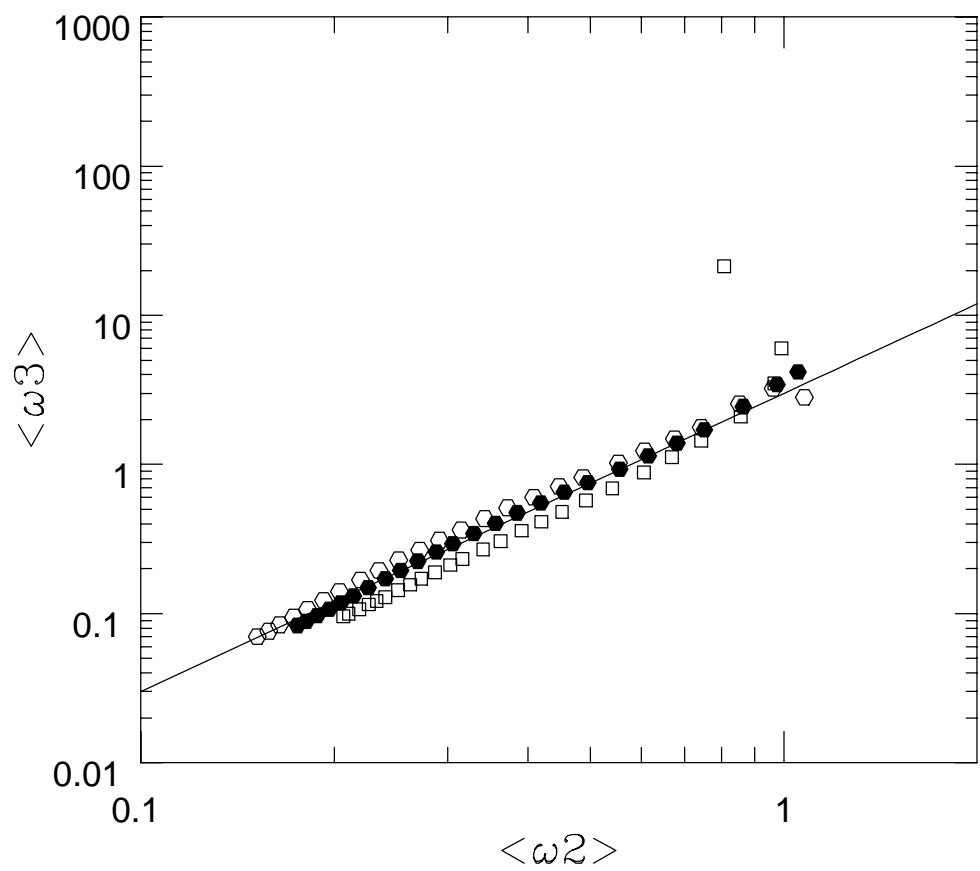

Figure 1b

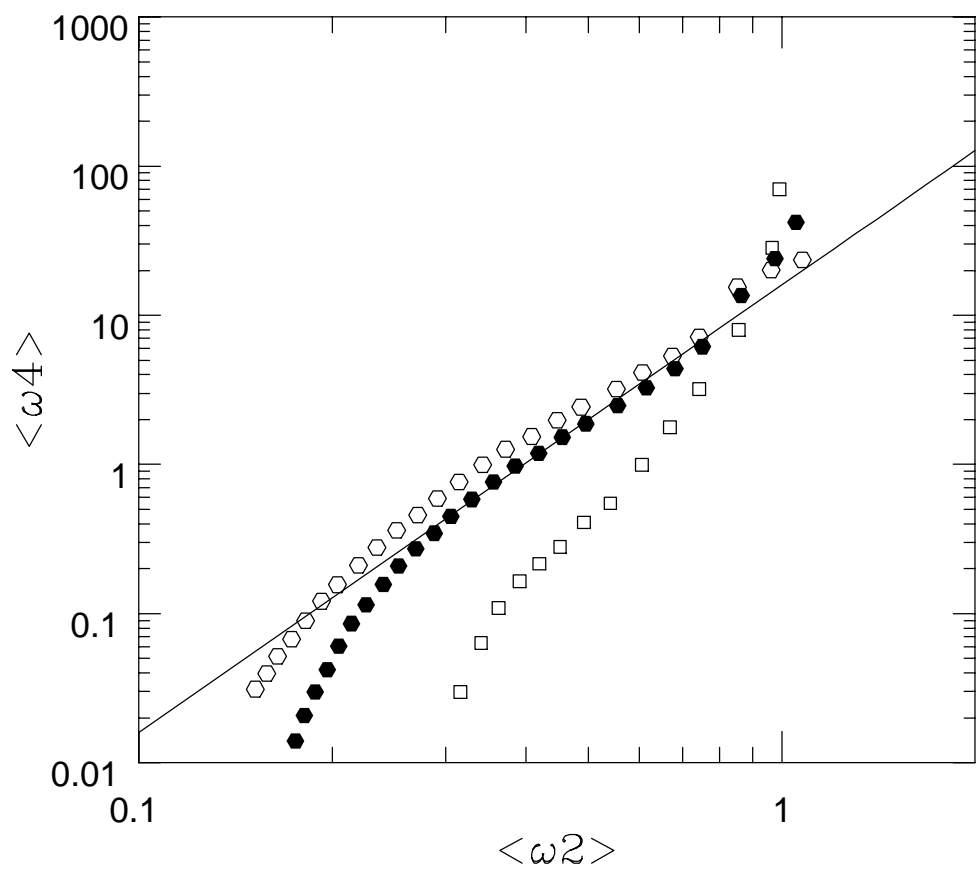

Figure 1c

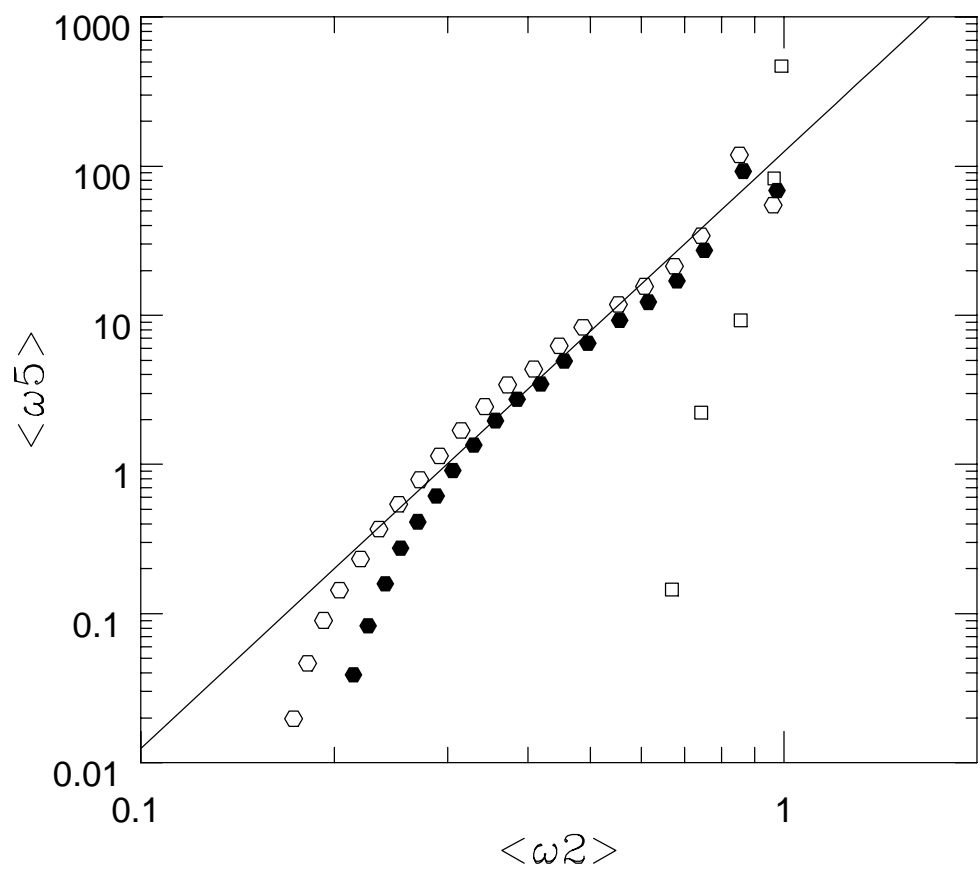

Figure 1d

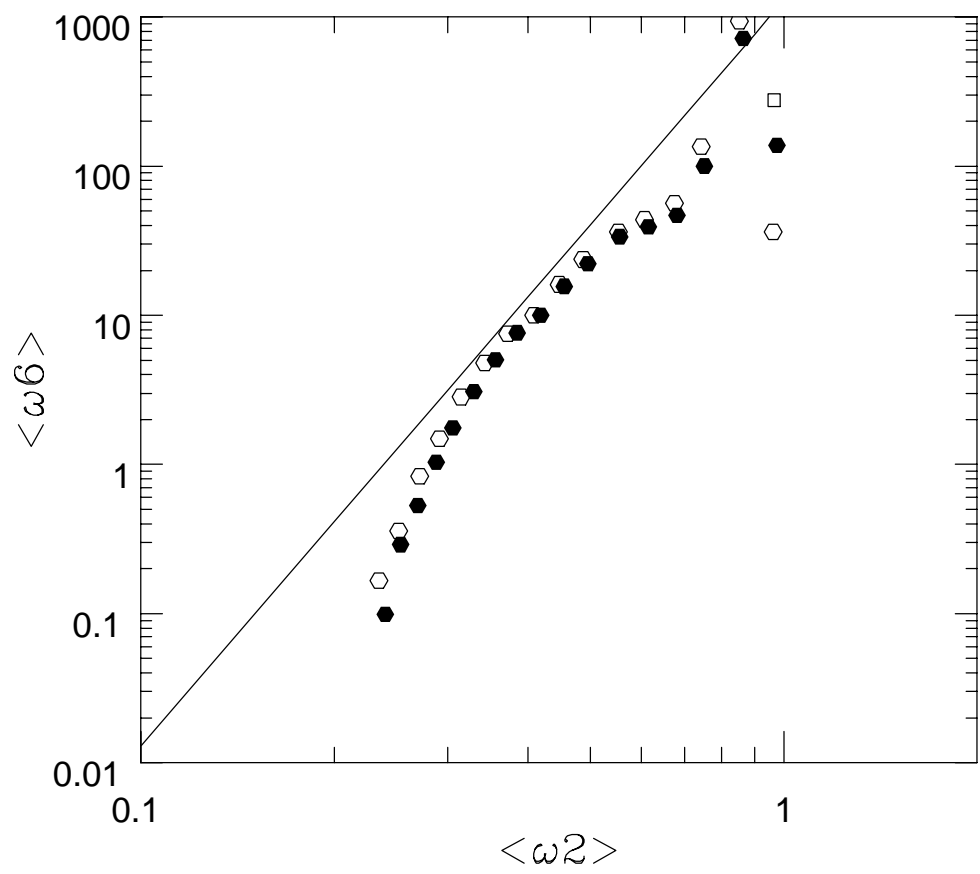

Figure 2

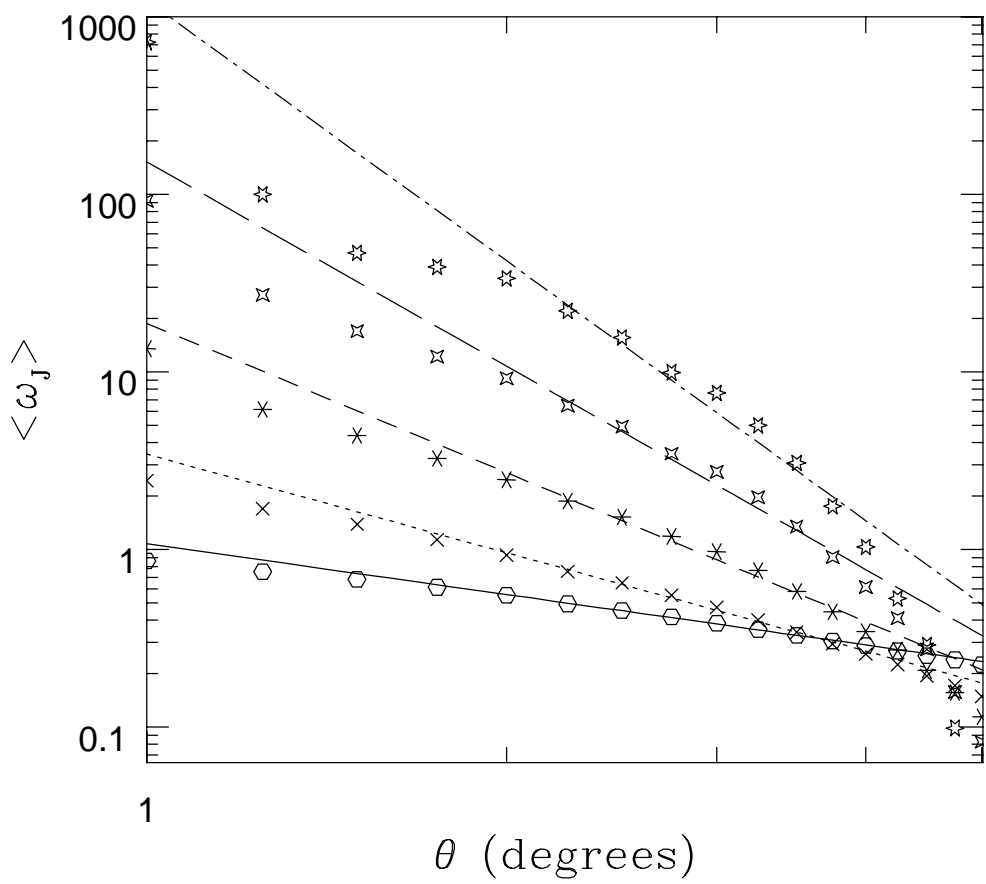

Figure 3

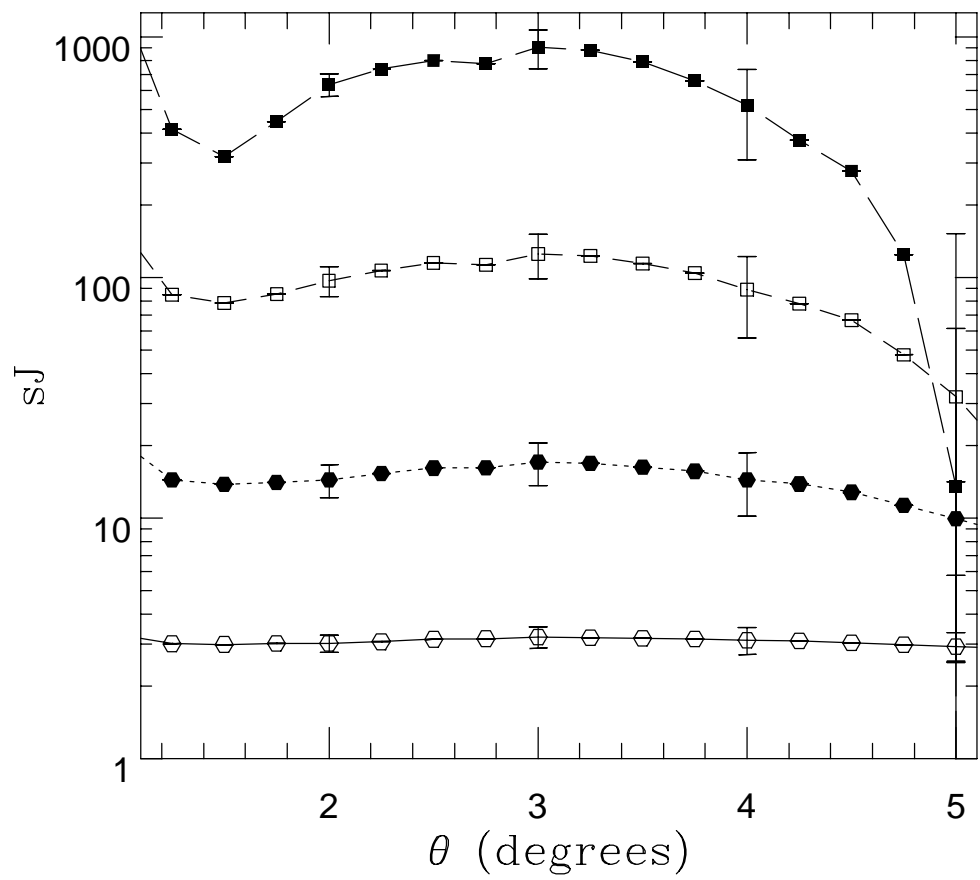

Figure 4

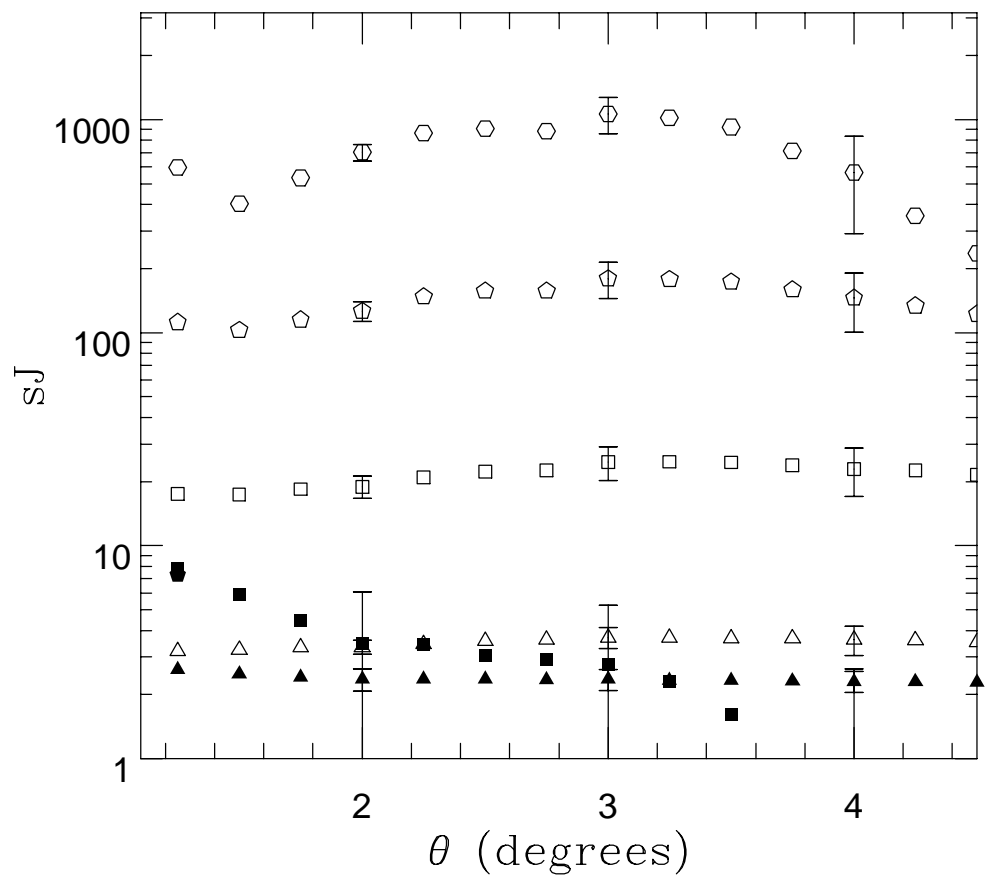

Figure 5

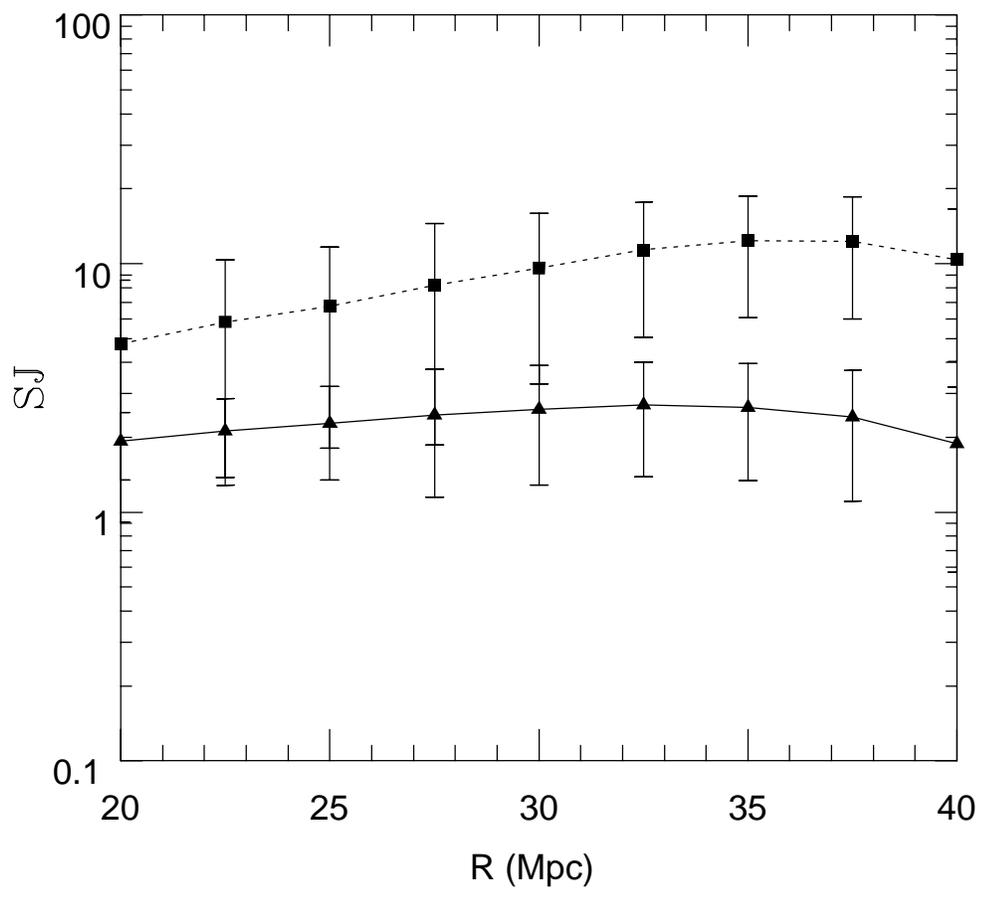